\documentclass[
 reprint,superscriptaddress,
 amsmath,amssymb,
 aps,
prb,
twocolumn
]{revtex4-2}
\usepackage{commath}
\usepackage{enumitem,kantlipsum}
\usepackage{tikz}
\usetikzlibrary{quantikz2}
\usepackage{graphicx}
\usepackage{dcolumn}
\usepackage{bm}
\usepackage{physics}
\usepackage{xcolor}
\usepackage[caption=false]{subfig}
\usepackage[normalem]{ulem}

\begin{document}
\title{Preparation of the single-spinon wave function on a quantum computer}

\author{D. Fa\'ilde}
\email{dfailde@cesga.es}
\affiliation{Galicia Supercomputing Center (CESGA), Avenida de Vigo, s/n, Santiago de Compostela, E-15782, A Coruña, Spain}
\author{A. Go\'mez}
\affiliation{Galicia Supercomputing Center (CESGA), Avenida de Vigo, s/n, Santiago de Compostela, E-15782, A Coruña, Spain}
\author{ J. Fernández-Rossier}
\affiliation{International Iberian Nanotechnology Laboratory (INL), Avenida Mestre José Veiga, 4715-330 Braga, Portugal
}

\begin{abstract}
We address the problem of how  to prepare 
single-spinon wave functions, relevant for one-dimensional   $S=1/2$ spin models,  in a quantum computer.  We adopt the recently proposed ansatz \cite{kulk} for the single-spinon wave function, where a state with $S=1/2$ is built in a spin chain with $L+1$ sites, adding a site with  $S_z=1/2$  to the configurations of the ground-state wave function for the spin chain with length $L$. We extend the original work to the case of the Haldane-Shastry model. We discuss how to prepare the single-spinon ansatz both for the Heisenberg and Haldane-Shastry models in quantum computers,  using a linear combination of unitaries.  We consider three different strategies to compute the single-spinon energy in a quantum computer, we analyze their cost in terms of the number of qubits, gates, and circuits, and we test them in-silico.

\end{abstract}

\maketitle

\section{Introduction}
The Heisenberg one-dimensional spin chain with $S=1/2$ with antiferromagnetic interactions has a quantum disordered ground state (GS) with $S=0$ (for 
even-numbered chains) and a gapless spectrum of spin excitations with $S=1$.  These excitations form a continuum in the ($k,E$) plane \cite{PhysRev.128.2131}, as opposed to magnon excitations with a well-defined energy-momentum relation, that occur for instance in systems with  long-range order in their ground states.

In 1981,  Fadeev and  Takhtajan~\cite{FADDEEV1981375} proposed that the physical spin excitations of the Heisenberg model are actually composite particles made of a much simpler object, dubbed as spinon, that has $S=1/2$ and a well-defined energy-momentum relation, given by $\varepsilon(q)= \pi/2\, cos(q) $. 
The idea was later extended to other spin models, such as the Haldane-Shastry model \cite{Haldane1,Shastry}, and the dispersion relation for spinons was found to be \cite{haldane91} $\varepsilon(q)=\frac{J}{2} (q_0^2-q^2)$, where $q_0=\frac{\pi}{2}$.
The concept of spinon has also played a role in the discussion of spin-charge separation of fermionic 1D models \cite{Luttinger,Haldane_1981,PhysRevLett.77.4054,Schofield}.

In the context of the Heisenberg model,  spinons do not exist as isolated objects,  the concept of a single-spinon wave function is not self-evident \cite{talstra97}. In this scenario, Ref. \cite{kulk} Kulka {\em et al.} have proposed a heuristic ansatz for the single-spinon wave function with a well-defined wave vector $q$,  spin $S_z=1/2$, that allows to compute their energy dispersion $E(q)$ for the XXZ model and correctly predicts their
 domain in the $q$ space, i.e, $-\pi/2 <q < \pi/2$.

In this work, we present two main contributions. First, we show that the single-spinon wave function proposed by Kulka {\em et al.}, also provides a good description of spinons in the Haldane-Shastry model. Second, and the main focus of this work,  we address the question of how to prepare the single-spinon ansatz proposed by Kulka {\em et al.}~\cite{kulk} in a quantum computer.  The proposed methods are only tested in-silico. The preparation of non-trivial spin wave functions in quantum computers has been widely studied in the last few years \cite{Cervera18,van21,Jasper22,murta23,Ciavarella2023,PRXQuantum.4.02031,Sierra24,Murta24,Tavernelli25,Cervera25, sym14030624}. This includes the preparation of the AKLT state \cite{murta23,PRXQuantum.4.02031}, the implementation of the Bethe ansatz \cite{van21,Sierra24, Raveh2024, Ruiz2025}, and the study of more intricate spin models that lack exact solutions, which can be explored using quantum algorithms like VQE \cite{sym14030624,Murta24,Jasper22}. Our main motivation is to propose a way to realize the single-spinon wave function in a physical system. A second motivation is to pave the way towards the implementation of spinon wave functions in two-dimensional models. There,  
the efficient techniques available in the one-dimensional case, such as Bethe Ansatz \cite{karbach98,Bethe} and Density Matrix Renormalization Group \cite{white92}, or its modern reformulation in terms of Matrix Product States \cite{schollwock11,orus14}, do not work or become inefficient in higher dimensions \cite{King}, respectively. In that situation, quantum simulation may offer the first practical route to access their wave functions and dynamics.

\begin{figure*}[t] 
    \centering 
    \includegraphics[width=1\textwidth]{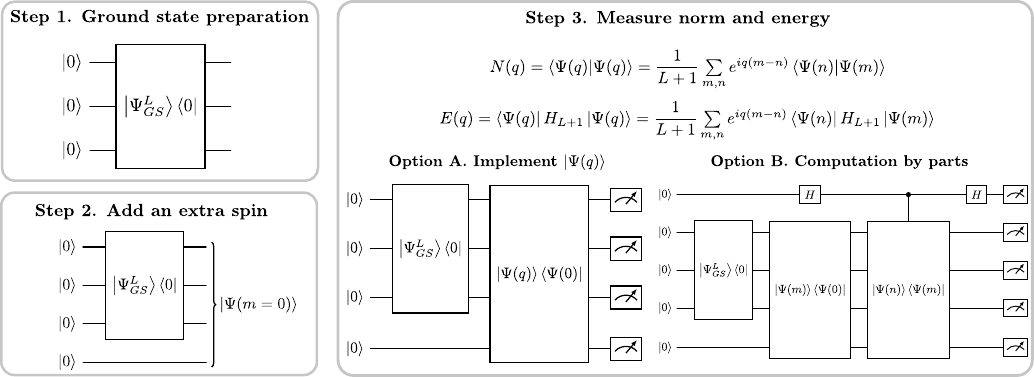} 
    \caption{Routine for the preparation of 1D single spinon states on a quantum computer. The protocol consists of three main steps. First, the ground state $\ket{\Psi_{GS}}$  of the target spin model is prepared on the quantum computer (step 1). Second, an additional spin is injected at site $m$ (step 2). Finally, we proceed to the evaluation of the norm $N(q)$ and energy $E(q)$. This can be achieved in two distinct ways: (option A) by implementing the single-spinon ansatz $\ket{\Psi(q)}$ and measuring directly both magnitudes quantumly; or by performing a part-by-part quantum computation where $N(q)$ and $E(q)$ are classically reconstructed (option B).      } 
    \label{fig:Scheme} 
\end{figure*}

The paper is structured as follows. In section \ref{sec:sp_ansatz}, we review the spinon ansatz of Kulka {\em et al.}~\cite{kulk} that permits computing the single spinon dispersion energy using the coefficients of the ground state wave function of the Heisenberg $S=1/2$ spin chain. In that section, we also show that the single-spinon ansatz also permits obtaining the spinon dispersion energy of the Haldane-Shastry Hamiltonian \cite{haldane91,Haldane2,Shastry}.
The preparation of the single-spinon state in a quantum computer requires, as a starting point, the preparation of the ground state of the spin model  (Fig.\,\ref{fig:Scheme}). This step is discussed in section \ref{sec:GSPREP}, for the case of two different Hamiltonians, the 1D Heisenberg and Haldane-Shastry models. In section \ref{sec:1spinon-QC}, we discuss the preparation of the one-spinon ansatz using a linear combination of unitaries. In section \ref{sec:norm-disp}, we discuss two deterministic methods that permit one to compute the energy dispersion of the spinons without preparing the state in the quantum computer.

\section{The single-spinon Ansatz \label{sec:sp_ansatz}}

\subsection{Review of the Ansatz}
Here, we review the single-spinon ansatz theory proposed by Kulka and co-workers \cite{kulk}.
The starting point is the wave function of the ground state of a given 1D  spin model for a chain with $L$ sites:
\begin{equation}
    \ket{\Psi_{GS}}=\sum_{j=0}^{2^L-1} A_j \ket{j}
    \label{eq:GS}
\end{equation}
 where $A_j$ are the amplitudes and $\ket{j}$ the elements of a given orthonormal basis:
\begin{equation}
\ket{j}=\ket{\sigma_0^j,\sigma_1^j,..\sigma_{L-1}^j}
\end{equation}
where $\sigma_n^j$ can take only two values, $\uparrow$ or $\downarrow$.
Since the ground state is a singlet, only the coefficients with total $S_z=0$ have a non-vanishing coefficient $A_j$.

In order to introduce  the  single-spinon ansatz, 
a set of extended configurations is introduced:
    \begin{equation}
\ket{j}=\ket{\sigma_0^j,\sigma_1^j,..\sigma_{L-1}^j}
\rightarrow \ket{j(m)} =
\ket{\sigma_0^j,\sigma_1^j,..,\uparrow,...\sigma_{L-1}^j}
\label{eq:jm}
\end{equation}
where the extra spin  $\uparrow$ is added after the first $m-1$ entries of $\ket{j}$. This describes a spin frozen at the $m$ site in a chain with $L+1$ spins, where the remaining spins preserve the correlations present in the GS. 
For that matter, the following state is introduced:
\begin{equation}
    \ket{\Psi(m)}=\sum_{j=0}^{2^L-1} A_j \ket{j(m)}.
\label{eq:localm}
\end{equation}
Since spinons have a well-defined wave-vector, the spinon-ansatz adopts the form \cite{kulk}:
\begin{equation}
    \ket{\Psi(q)}= \frac{1}{\sqrt{L+1}} \sum_{m=0}^{L} e^{iqm} \ket{\Psi(m)}.
\label{Fourier}
\end{equation}

The usefulness of this ansatz comes from the fact
that the 
energy dispersion $\epsilon(q)$ with respect to the GS energy can be obtained as
\begin{equation}
\epsilon(q)=\frac{\bra{\Psi(q)}H_{L+1}\ket{\Psi(q)}}{N(q)} -E_0^{L+1},
\label{eq:spinon-disp}
\end{equation}
being $E_0^{L+1}$ the GS energy of the chain with $L+1$ sites, and
\begin{equation}
    N(q)=\bra{\Psi(q)}\ket{\Psi(q)}
    \label{Norm}
\end{equation}
the {\em momentum dependent norm}. The resulting energy exhibits two key properties. First, it compares well with the analytical results \cite{kulk}. Second, the value of the norm $N(q)$ goes to zero, as $L$ increases, out of the
$-\pi/2  \leq q \leq \pi/2$ region, giving the momentum domain in which spinons are defined. 
Therefore, the spinon ansatz captures well the main features of spinons as elementary quasiparticles: their energy, momentum, and spin.

\subsection{Application of the single-spinon ansatz to the Haldane-Shastry model}
In the original paper of Kulka {\em et al.}~\cite{kulk}, the energy dispersion of single spinons for the first-neighbour XXZ model was calculated, and compared successfully with analytical results \cite{FADDEEV1981375}. 
 Here, we test  the validity of the single-spinon ansatz to the case of the Haldane-Shastry model \cite{haldane91,Haldane2,Shastry}, which describes a $S=1/2$ spin chain with long-range  isotropic exchange couplings:
\begin{equation}
    H_{HS}=\frac{J\pi^2}{L^2}\sum_{n<m} \frac{\vec{S}_n \cdot \vec{S}_m}{\sin^2(\pi(n-m)/L)}.  
\end{equation}

This model is exactly solvable and, as in the case of the Heisenberg chains, the ground state has $S=0$, and its physical $S=1$ excitations are composite particles of spinons that have a well-defined energy-momentum relation and $S=1/2$ \cite{Haldane1}. Its ground state can be expressed as the Gutzwiller projected Fermi sea for free fermions (with single-particle states $e^{ik n}$, where $n$ labels the chain site). This fact becomes useful later, when it comes to preparing the single-spinon ansatz in a quantum computer, as there are quantum algorithms to prepare the Gutzwiller state \cite{Murta21}.

\begin{figure}[t] 
    \centering 
    \hspace{-1cm}
    \includegraphics[width=0.48\textwidth]{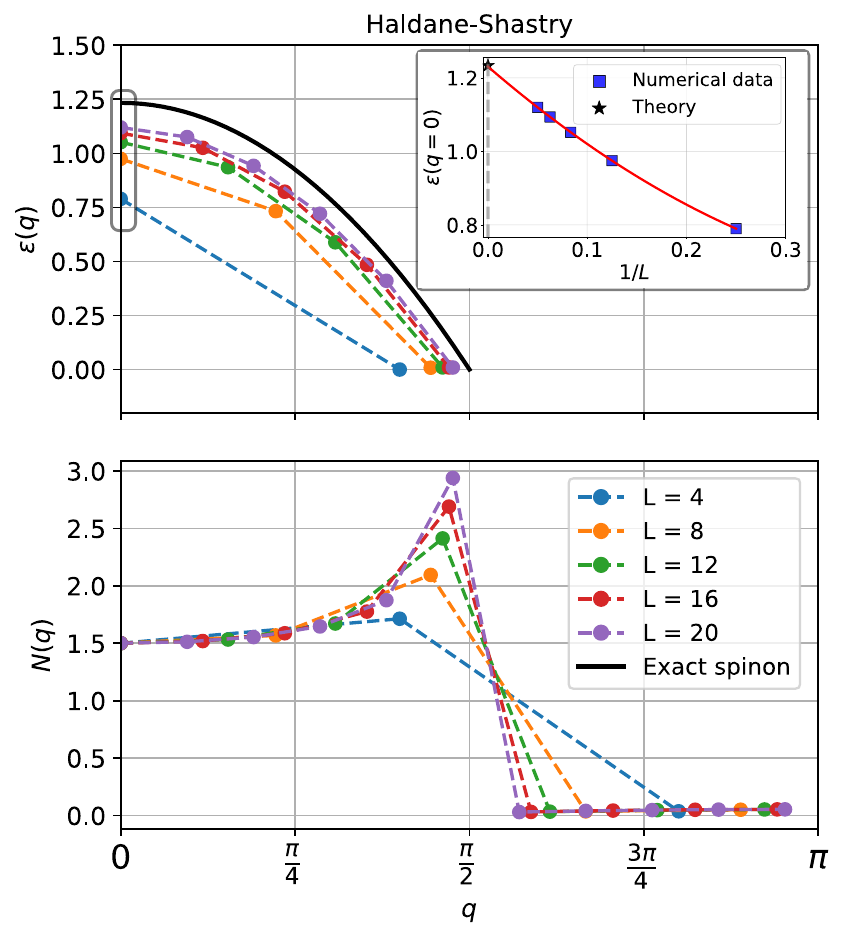} 
    \caption{(a) Energy dispersion $\epsilon(q)$ and (b) norm $N(q)$ of the single-spinon wavefunction $\ket{\Psi(q)}$ for the antiferromagnetic Haldane-Shastry model, calculated in-silico. The norm drastically decreases for momenta $q>\pi/2$ while the energy dispersion $\epsilon(q)$ approaches the theoretical form as $L$ increases. In the inset, we show the energy $\epsilon(q=0)$ as a function of $1/L$. The red line represents a second-order polynomial fit, $f(1/L)=a+b/L+c/L^2$, to the numerical data (blue squares) extracted from the main plot. The starred point indicates the theoretical value of the energy at zero momentum in the thermodynamic limit, $\epsilon(q=0)=\pi^2/8$, taking $J=1$.} 
    \label{fig:HS} 
\end{figure}

In order to test the single-spinon ansatz in the context of the Haldane-Shastry model, we first find the ground state by numerical diagonalization of the spin model in rings with $L$ sites, with $L$ up to 20  sites. Without loss of generality, we can take $J=1$, as this energy scale is a prefactor of the Hamiltonian operator. In Fig.\,\ref{fig:HS}a we show the estimation of the single-spinon energy dispersion obtained using  Eq. (\ref{eq:spinon-disp}) along with the analytical formula derived by Haldane \cite{haldane91} $\varepsilon(q)=\frac{J}{2} (q_0^2-q^2)$. In the inset, we show $\epsilon(q=0)$, estimated with this method, as a function of $1/L$. The red line is a fit, computed without adding the $q=0$  analytical value. We find that the fitting curve extrapolates well in the limit of large $L$, approaching the analytical calculation. 

In Fig.\,\ref{fig:HS}b, we compute the norm of the single-spinon ansatz (Eq. (\ref{Fourier})) for the Haldane Shastry model.  As the value of $L$ is increased, it becomes apparent that the norm vanishes for $q>\frac{\pi}{2}$, in agreement with the fact that spinons are only supported in half of the Brillouin zone \cite{FADDEEV1981375}.
From our calculations, we can conclude that the single-spinon ansatz works as well in the case of the Haldane-Shastry model. 

\section{Ground State Preparation on Quantum Computers \label{sec:GSPREP}}

The strategy to prepare the single-spinon ansatz for a given model necessarily starts by preparing the ground state (GS)  in the quantum computer (Fig.\,\ref{fig:Scheme}), as the GS wave function coefficients, defined in Eq.\,\eqref{eq:GS} enter the single-spinon ansatz of Eq.\,\eqref{Fourier}. In a second stage, an ancilla qubit has to be added, to extend the 
lattice site as prescribed in equations \eqref{eq:jm} and \eqref{eq:localm}, followed by the Fourier transform of Eq.\,\eqref{Fourier}. In this section, we discuss the first step, namely, the preparation of the ground state wave function.
We consider two different models for 1D $S=1/2$ spin chains, the first-neighbour Heisenberg model and the Haldane-Shastry model.

In the case of the Heisenberg model, the strategy we choose to prepare the ground state of the model in a quantum computer is the so-called \emph{Variational Quantum Eigensolver} (VQE) \cite{Peruzzo2014}. This offers a flexible and widely used method \cite{TILLY20221}. It approximates the ground state using parametrized quantum circuits optimized via classical feedback, enabling the study of broader spin systems beyond exactly solvable models, with potential extensions to higher dimensions. While VQE is generally less computationally demanding than other quantum algorithms, such as Quantum Phase Estimation \cite{Lloyd99} or Quantum Imaginary Time Evolution \cite{Motta2020},  it provides only an approximate solution and is also subject to limitations. The performance of VQE is often hindered by the presence of local minima in the optimization landscape and the emergence of barren plateaus \cite{Larocca2025}, which severely limit the trainability and pose significant challenges for scaling the method to larger systems.

\begin{figure*}[t]  
  \centering
  \includegraphics[width=\textwidth]{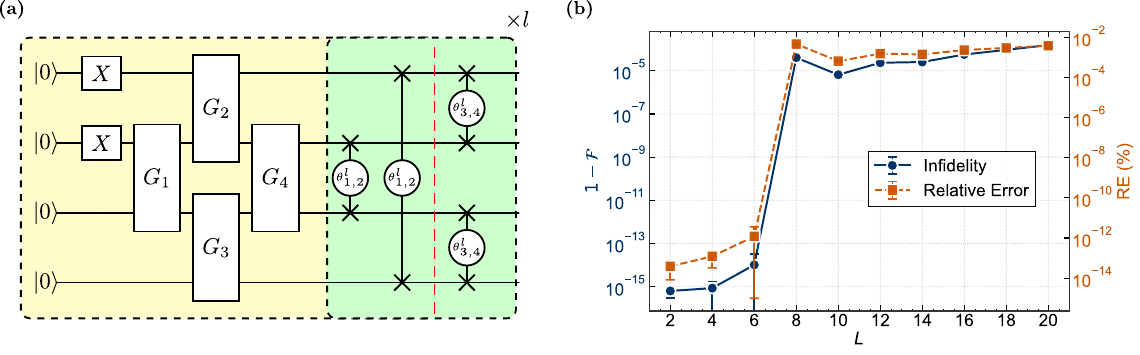} 
  \caption{
    (a) Illustration of the PQC used to prepare the ground state (GS) of the 1D Heisenberg XXX model for $L=4$ spins. 
    The yellow box represents the free-fermion state obtained via Givens rotations, while the green box corresponds to the parametrized ansatz. The red dashed line separates the parametrized gates generated by the two commuting groups of terms in the Heisenberg Hamiltonian. 
    (b) Infidelity $1-|\bra{\Psi_\theta}\ket{\Psi_{exact}}|^2$ and relative error $|\frac{E_\theta - E_0^L}{E_0^L}| $ as a function of the number of spins in the chain, $L$. The results indicate that the ansatz is sufficiently expressive to prepare the ground state with low infidelity and energy error, but, beyond $L = 4$, it no longer achieves floating-point-level accuracy. 
  }
  \label{fig:infidelity_error}
\end{figure*}

For the Haldane-Shastry chain \cite{Haldane1,Shastry,ANDERSON1973153}, we design a tailored quantum circuit to prepare its ground state in an exact way, taking advantage of two known facts. First, the ground state of the Haldane-Shastry model is given by the Gutzwiller projection \cite{Haldane1, Murta21} of a free fermionic model with nearest-neighbour interactions. Second, we build upon previous work proposing algorithms to carry out the Gutzwiller projection \cite{Murta21,Yunoki22,kang2026scalablequantumalgorithmsgutzwiller}.

We thus present two complementary routes to ground-state preparation. We note that, as an alternative to the methods used here, it would be possible to exploit the fact that the Heisenberg chain can be solved with the Bethe ansatz \cite{Bethe1931}, and prepare Bethe-ansatz states using recently proposed quantum algorithms \cite{van21,Sierra24, Raveh2024, Ruiz2025}. While some of these methods are non-deterministic, others achieve deterministic state preparation at the cost of exponential classical pre-processing. In either case, the overall resource cost grows rapidly with system size, limiting scalability for large $L$. A VQE approach can also be applied to the Haldane-Shastry Hamiltonian, though it also faces  challenges arising from its long-range, all-to-all interactions (see Appendix~\ref{B}).

\subsection{Preparation of ground state for first neighbour Heisenberg chain\label{subsec:1DHEIS}}
To prepare the GS of the 1D Heisenberg antiferromagnet with first neighbour interactions
\begin{equation}
    H=J \sum_n \vec{S}_n \cdot \vec{S}_{n+1},
\end{equation}
and periodic boundary conditions, we follow the shallow variational protocol recently used by one of us \cite{Murta24} that includes two steps.  First, the initialization,  where the quantum computer is loaded with  a valence bond crystal (VBC) (i.e., a state where every pair of adjacent qubits is prepared in a singlet)
\begin{equation}
    \ket{\Psi_{0}}=\frac{1}{\sqrt{2^{L/2}}} (\ket{\uparrow \downarrow}-\ket{\downarrow\uparrow} )^{\otimes L/2}.
\end{equation}
In the second step, an engineered parametrized set of unitaries is applied on the initial state, similar to a Hamiltonian Variational Ansatz (HVA) \cite{PRXQuantum.1.020319, Park2024hamiltonian}. Although accurate results can be achieved with a few layers in the ansatz,  increasing the number of variational parameters poses a significant challenge when scaling the method to larger system sizes.  Here, we have explored an alternative for the first step,  using instead of the VBC  state 
the GS of $H_{XY}=-J\sum_i S_i^x S_{i+1}^x+S_i^y S_{i+1}^y$. The $H_{XY}$ Hamiltonian maps to a spinless 1D free fermion model~\cite{lieb61}. At half-filling, we can construct the $H_{XY}$ ground state through Givens rotations \cite{Wecker15,Jiang18}. Fig.\,\ref{fig:fidelidad} illustrates the overlap between the Heisenberg GS $\ket{\Psi_{GS}}$ and the initial state $\ket{\Psi_0}$ for both the VBC and free fermionic states. The results show that the free fermionic state remains closer to the true ground state as $L$ increases, with the overlap decaying approximately linearly with  $L$ over the range considered. 
The enhanced overlap of the $H_{XY}$ ground state comes with the overhead in preparation, as  it 
requires $O(L^2)$ Givens rotations and depth $O(L)$ \cite{Wecker15,Jiang18}, to be compared with the one-shot preparation of the product of singlets. 

\begin{figure}[h]   
    \centering
    \includegraphics[width=0.8\columnwidth]{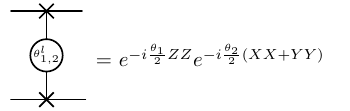}
    \caption{Parameterized gate for the Heisenberg interaction.}
    \label{gate}
\end{figure}

For the parametrized quantum circuit (PQC), we adopt the structure introduced in Ref.~\cite{Murta24}, consisting of blocks of $R_{ZZ}$ gates followed by $R_{XX+YY}$ gates, as illustrated in Figure~\ref{fig:infidelity_error}a. These gates come from the decomposition of the Heisenberg Pauli terms into unitaries (Fig.~\ref{gate}). Each block of non-commuting Pauli groups is associated with the same set of variational parameters. Therefore, in a circuit of $l$ layers, we have $4l$ parameters.

\begin{figure*}[t]  
  \centering
  \includegraphics[width=\textwidth]{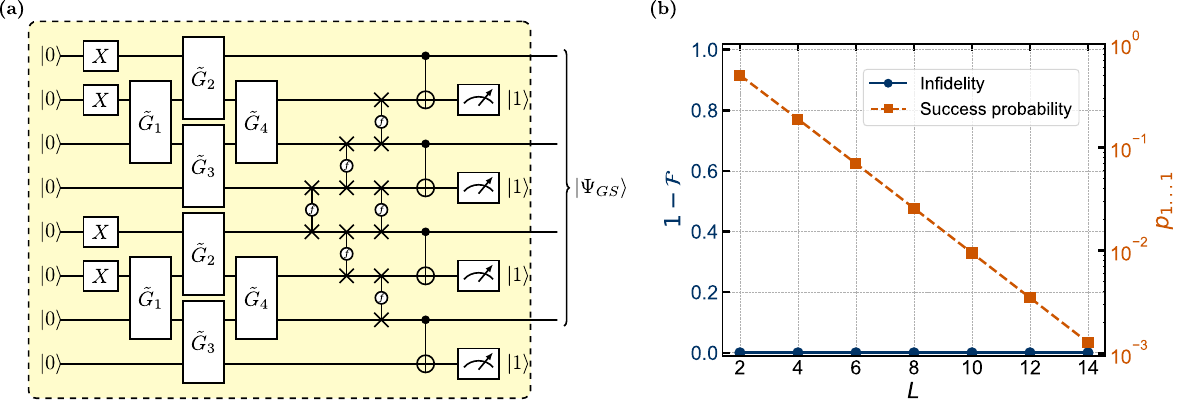}  
  \caption{
    (a) Quantum circuit that prepares the ground state (GS) of the 1D  Haldane-Shastry model for $L=4$ spins, using the Gutzwiller projection. SWAP gates labeled with $f$ denote fermionic SWAPs, introducing the correct phase when the two fermionic modes are occupied \cite{Murta21}.
    (b) Infidelity and probability of implementing the GS as a function of the number of spins in the chain, $L$. Fidelity remains maximal under post-selection of $\ket{1}$  outcomes, but the success probability decays exponentially with system size. To accurately estimate the success probability $p_{1...1}$, we executed the circuit with $10^8$ shots. 
}
  \label{fig:haldane-shastry}
\end{figure*}

As shown in Figure~\ref{fig:infidelity_error}b, by setting $l = L/2$, we can maintain a consistently low infidelity across system sizes, keeping it well below the threshold expected from sampling on real quantum devices. This is achieved using a PQC of moderate depth $O(L)$, with the total number of parameters scaling linearly as $N_\theta = 2L$, which is well-suited for near-term quantum hardware and facilitates trainability during the optimization process. The results highlight the effectiveness of this architecture in accurately approximating the ground state of medium-length Heisenberg XXX chains on quantum computers. Similar performance is expected for the anisotropic Heisenberg XXZ model in the regime $J_z < J_{xy}$. All results in Fig.\,\ref{fig:infidelity_error} were obtained after ten different optimizations per data point using the default \texttt{SLSQP} algorithm from SciPy. The initial parameters of the circuit were set according to the procedure outlined in Ref.~\cite{Park2024hamiltonian}.

\subsection{Preparation of the Haldane-Shastry ground state \label{subsec:HS}}

For the preparation of the ground state wave function of the  1D Haldane-Shastry model, we use the fact
that this is given by the Gutzwiller projected wave function for tight-binding fermions in the ring. This wave function is derived in two steps \cite{Murta21}. First, we build the ground state of the free fermion Hamiltonian

\begin{equation}
   H= -t \sum_{i,\sigma}^L c_{i,\sigma}^\dagger  c_{i+1,\sigma} + h.c
    \label{free_fermions}
\end{equation}
where $t$ is the hopping amplitude. The ground state of \eqref{free_fermions} corresponds to a filled Fermi sea (FS) 
\begin{equation}
    \ket{\text{FS}} = \prod_{|k| \leq k_F,\sigma} c_{k,\sigma}^\dagger \ket{0}=\frac{1}{\sqrt{L}} \prod_{|k| \leq k_F,\sigma}\left(\sum_{j}^L e^{ikj} c_{j,\sigma}^\dagger\right) \ket{0}
\end{equation}
where $ c_{k,\sigma}^\dagger = 1/\sqrt{L} \sum_{j} e^{ikj} c_{j,\sigma}^\dagger $ creates a fermion with momentum $k$ and spin $\sigma$. The Fermi momentum $ k_F = \pi/2 $ guarantees that there is an electron per site on average. This free-fermion state can again be efficiently prepared on a quantum computer using Givens rotations \cite{somma02,Wecker15,Jiang18,Murta21}, which implement the transformation from the momentum-space to the real-space.

In a second step, a Gutzwiller projection $P_G=\prod_i^L(1-n_{i,\uparrow}n_{i,\downarrow})$  is applied on the site basis,  eliminating the configurations where a site is occupied for both spins. It can be implemented via a sequence of $\rm CNOT$ gates between spin-$\uparrow$ and spin-$\downarrow$ subsystems (Figure\,\ref{fig:haldane-shastry}). The operation is implemented upon obtaining $\ket{1}$ in measurements of all $\downarrow$-spin sites, enforcing single occupancy. As a non-unitary transformation, the implementation of $P_G$, and hence the preparation of the GS of the 1D Haldane-Shastry model, leads to a non-deterministic routine whose success probability decays exponentially with the number of qubits. Numerical fitting reveals that the probability $p_{1\ldots1}$ of obtaining all $\ket{1}$ outcomes scales as $2^{-\sqrt{2}\, L / 2}$,  limiting its practical application to moderate-length chains.

A VQE approach for the Haldane-Shastry model is also possible, though more complex than for the Heisenberg chain. The ground state of $H_{XY}$, $\ket{\Psi_0^{XY}}$, remains a suitable warm-start for medium-length chains, as it has significant overlap with the true ground state, as shown in Figure~\ref{fig:FF}. However, beyond optimization challenges, implementing the PQC from Figure~\ref{fig:infidelity_error}a for $H_{\text{HS}}$ results in a circuit whose depth per layer grows with $L$. Moreover, the all-to-all interactions require high qubit connectivity, necessitating a growing number of SWAP gates on limited-topology hardware, further increasing depth and noise sensitivity.

\section{Preparation of the single-spinon wave function on a quantum computer\label{sec:1spinon-QC}}
\subsection{Overview of the method}
The preparation of the spinon ansatz wave function $\ket{\Psi(q)}$ is carried out using the following steps.  First,  the preparation of the state $\ket{\Psi(m=0)}$ of Eq.\,\eqref{eq:localm} in a digital quantum processor is carried out trivially, by adding an ancilla qubit in the state $|0\rangle=\uparrow$ at one end of the chain.

Second,  the preparation of the state $e^{i q m} \ket{\Psi(m)}$ is obtained by the repeated application  of a modified  SWAP unitary operator to $\ket{\psi(m=0)}$ 
\begin{equation}
    U(q) = e^{iq}
\begin{pmatrix}
1 & 0 & 0 & 0 \\
0 & 0 & 1 & 0 \\
0 & 1 & 0 & 0 \\
0 & 0 & 0 & 1
\end{pmatrix}.
\label{eq:U}
\end{equation}

The third step, the preparation of the state \eqref{Fourier}, is clearly a linear combination of unitaries (LCU). In turn, this relates to the facts that the single-spinon ansatz is not normalized, and there is no single unitary operator that can map $\ket{\psi(m=0)}$ into Eq.\,\eqref{Fourier}.
To illustrate this,  consider the case where $\ket{\Psi_{\mathrm{GS}}} = \ket{\uparrow\downarrow}$ for $L=2$. Adding an ancillary spin $\uparrow$ at position $m=0$ yields the state $\ket{\Psi(0)} = \ket{\uparrow \uparrow \downarrow}$. Applying the Fourier transform \eqref{Fourier}, we obtain the state
\begin{equation}
    \ket{\Psi(q)}=\frac{1}{\sqrt{3}}\left( \ket{\uparrow\uparrow \downarrow}(1+e^{iq})+\ket{\uparrow\downarrow\uparrow}e^{i2q} \right)
\end{equation}
where the swaps that involve the same spin are the ones that cause the non-unitarity, $\ket{\Psi(0)}=\ket{\Psi(1)}$. Moreover,  it is important to emphasize that, for the Heisenberg and Haldane-Shastry models, $\ket{\Psi_{GS}}$ is definitely not a simple product state but an entangled state. Consequently, the number of required non-unitary operations is, in general, $L$, which stands for the number of spin translations $m=1,...,L$ applied to the input state $\ket{\Psi(0)}$.

\begin{figure*}[t]
\includegraphics[scale=0.85]{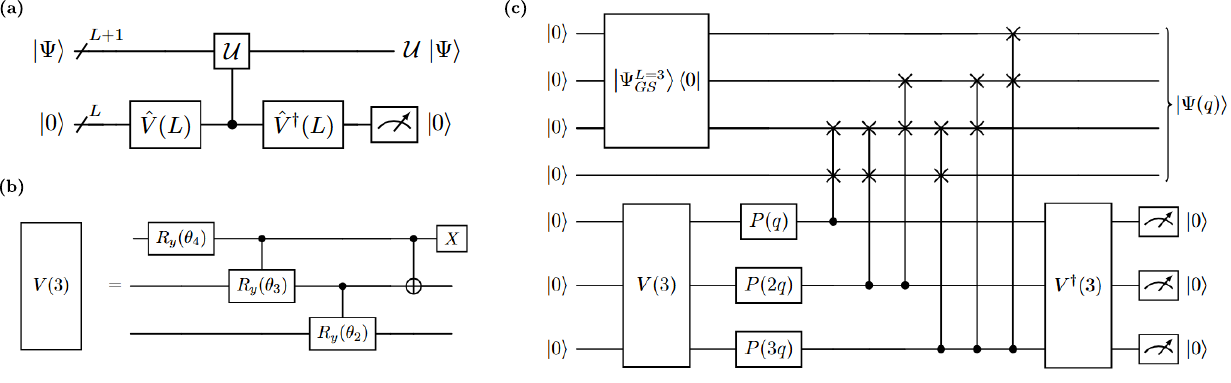}~~~~
\caption{
(a) High-level scheme of the implementation of the single-spinon ansatz. Given an input state $\ket{\Psi}$ on $L+1$ qubits, the $L$ ancilla qubits, prepared in the state $\ket{V(L)}$, conditionally apply the operator $\mathcal{U}$ to the main registers. 
The LCU protocol is successfully carried out if the measurement of the ancilla output yields $\ket{0}^{\otimes L}$. (b) Scheme to prepare the state $\ket{V(L)}=\hat{V}_L \ket{0}^{\otimes L}$ in the ancillary registers, illustrated for $L=3$ qubits. For $L$ ancillas, the circuit consist on $L$ parametrized $R_y$ gates, being $\theta_k=2arccos(1/\sqrt{k})$. The number of controlled operations is $2L-3$. (c)   Quantum circuit for the direct implementation of the momentum eigenstate $\ket{\Psi(q)}$ on a quantum computer using Linear Combinations of Unitaries, illustrated for a chain of $L=3$ spins. The input state consists of the ground state $\ket{\Psi_{\mathrm{GS}}^L}$ of the model, augmented by an additional spin $\uparrow$. An ancillary register prepared in the state $\ket{V(L)}$ enables the application of the operator $\mathcal{U}$ that implements the Fourier Transform. The target state $\ket{\Psi(q)}$ is probabilistically obtained upon measuring the ancillary qubits and post-selecting the outcome $0$. }
\label{fig:LCU}
\end{figure*}

\subsection{LCU routine}
To implement in a quantum processor the full operator that produces Eq.\,\eqref{Fourier}, we make use of the algorithm \cite{Childs12,murta23} to prepare a LCU,
which entails the use of ancilla qubits and a non-deterministic approach. The LCU operator is expressed as

\begin{equation}
    \mathcal{U}=\sum_{m=0}^L U_m
    \label{eq:LCU}
\end{equation}
where $U_0=I$, and 
\begin{equation}
  U_m=e^{imq} {\rm SWAP}_0^m  
  \label{eq:Um},
\end{equation}
being ${\rm SWAP}_0^m={\rm SWAP}_{m-1}^m...{\rm SWAP}_0^1$  a ladder of $m$ SWAP gates which brings the spin from the position $0$ to $m$. For clarity, each unitary operator $U_m$ in \eqref{eq:LCU} transforms the input state $\ket{\Psi(0)}$ in $e^{iqm}\ket{\Psi(m)}$. Therefore, 
the operator Eq.\,\eqref{eq:LCU} 
maps the state of Eq.\,\eqref{eq:localm}, with $m=0$, into  the state of Eq.\,\eqref{Fourier}.

The implementation of the LCU is carried out through a non-deterministic algorithm \cite{Childs12,Chakraborty2024implementingany} that introduces $L+1$ ancilla qubits, one per term in the sum of eq. \eqref{eq:LCU}. In our case, we do it using $L$ ancillary qubits as $U_0$ corresponds to the identity. The number of ancillary qubits can be reduced up to $n\geq log_2(L+1)$ using all the possible states in $n$ qubits. However, this requires the usage of multi-controlled operations instead of single controls, besides the difficulty of encoding the relative phase $e^{iq}$. 

\begin{figure*}[t]
\includegraphics[scale=0.9]{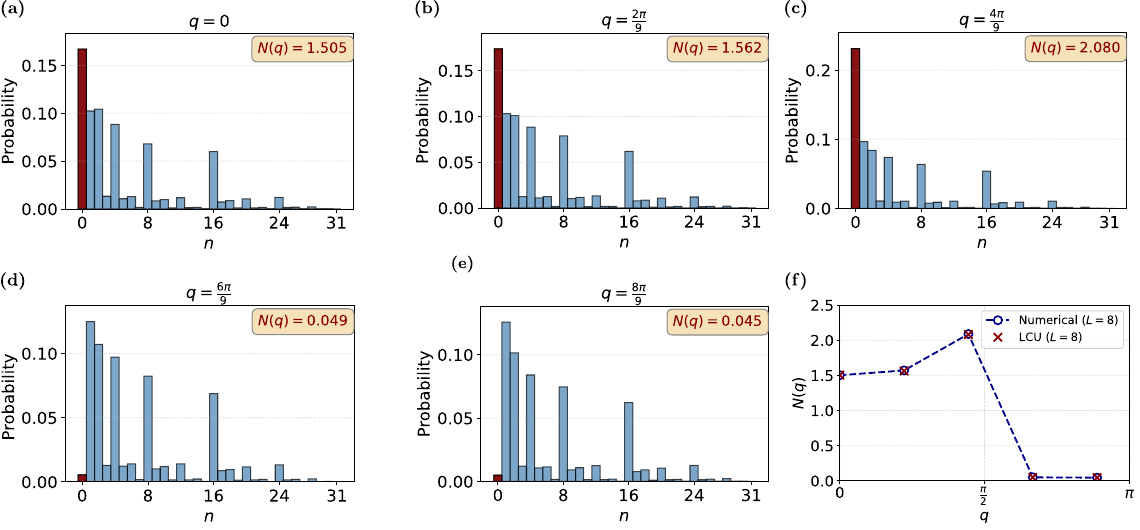}~~~~
\caption{Norm estimation of the single spinon wave function prepared via LCU for an $L=8$ Heisenberg model.  (a-e) Probability distribution truncated to the first $32$ computational basis states of the ancillary qubits, obtained via simulation of the LCU circuit with $10^5$ shots. The probability of measuring the all-zero state $\ket{0}^{\otimes L}$ (highlighted in red) is used to estimate the norm $N(q)$ for different values of $q \in [0, \pi]$.  As the input of the LCU protocol, the ground state of the Heisenberg model was prepared using the PQC introduced in the previous section, with the optimized parameters obtained with the VQE algorithm. (f) A comparison of the norm $N(q)$ obtained through exact diagonalization and estimated via the LCU protocol with $10^5$ shots for a Heisenberg chain with $L=8$ spins.}
\label{fig:LCU_results}
\end{figure*}

The overall structure of the LCU algorithm, which we follow here, is shown in Fig.\,\ref{fig:LCU}(a). 
In the first stage, a unitary operator $V(L)$ is applied over the ancilla qubits, initially all at $\ket{0}$, i.e., ${V}(L)|0\rangle^{\otimes L}\equiv \ket{V(L)}$, resulting in the   state:
\begin{equation}
  \ket{V(L)}=\frac{1}{\sqrt{L+1}} \left(\sum_i^{2^L} \ket{i} \right)_{Hamming(i)\leq 1}
    \label{eq:VL}.
\end{equation}
The state \eqref{eq:VL} represents an equal superposition of states with Hamming weights less than or equal to 1. For instance, $\ket{V(2)}=\frac{1}{\sqrt{3}}(\ket{00}+\ket{{01}}+\ket{10})$. This state can be generated through a downwards ladder of controlled $R_y(\theta)$ gates followed by a ladder of $\rm CNOTs$   (Figure\,\ref{fig:LCU}(b)). Each ancilla qubit $m$ controls the operator $U_m$, given by equation \eqref{eq:Um}.
To complete the LCU protocol, we apply the inverse operator $V^\dagger(L)$ and measure the ancillary registers. The momentum eigenstate $\ket{\Psi(q)}$  only is  
implemented whenever the readout of all the ancillary qubits is zero 
(Figure\,\ref{fig:LCU}c).

\subsection{Estimation of $N(q)$ and $E(q)$ using the single-spinon ansatz}
The probability of obtaining $\ket{0}^{\otimes L}$ in the ancillary qubits is given by $\frac{N(q)}{L+1}$. This contrasts with the exponential decay in the success probability of the non-deterministic preparation of the Haldane-Shastry ground state. In this case, a linear decay renders the LCU approach feasible for larger system sizes. Therefore, if we execute the LCU circuits   $n_{\text{shots}}$ times, for a given value of $q$ and $L$, and we find $n_0(q)$ times all the ancilla qubits in the $0$ state (see Fig.~\ref{fig:LCU_results}a--f), we 
obtain an estimate for the norm of the single-spinon wave function given by: 
\begin{equation}
    N(q) = \frac{n_0(q)}{n_{\text{shots}}}(L+1)
\end{equation}
In Fig.\, \ref{fig:LCU_results} we illustrate the calculation of the norm for different single-spinon states, with different momenta $q=\frac{2\pi n}{L+1}$, for the Heisenberg model with  $L=8$. The ground state has been obtained using VQE.  For each $q$ value, the norm is obtained after $10^5$ shots. Results demonstrate strong quantitative agreement over the full range of $q \in [0, \pi]$.  In Fig.\,\ref{fig:LCU_results}a--e, we show the histograms of the readout probabilities for the ancilla states, for the different values of $q$.  For clarity, we only show the first $2^5$ computational states.


Once the single-spinon state of Eq.\,\eqref{Fourier} is prepared, the expectation value $\bra{\Psi(q)}H_{L+1}\ket{\Psi(q)}$ can be computed by doing tomography for the different terms in the Hamiltonian. This will require   
rotating the other $L+1$ qubits to the appropriate measurement basis for each Pauli term or commuting group in $H_{L+1}$ \cite{TILLY20221}. 
On the other hand, implementing $\ket{\Psi(q)}$ requires a total of $\frac{L(L+1)}{2}$ Fredkin gates per circuit, assuming all-to-all qubit connectivity. This count can be reduced to $\frac{L(L/2+1)}{2}$ Fredkin gates by initially placing the extra spin in the middle of a chain with an even number of spins, under the same connectivity assumptions. Nevertheless, current quantum hardware remains limited in its ability to scale such circuits to large system sizes. Trapped-ion devices, with their inherent all-to-all connectivity, provide a favorable architecture for these tasks~\cite{Wright2019}, although alternative platforms are actively evolving to overcome similar challenges. For instance, native three-qubit gates have been proposed on both trapped-ion platforms via spin-dependent squeezing~\cite{PhysRevLett.129.063603} and superconducting processors with tunable couplers~\cite{lvb9-pfr3}. If native 3-qubit gates are not available, each Fredkin gate can be decomposed into five two-qubit gates or, when restricted to CNOTs only, into 8 CNOT gates~\cite{PhysRevA.53.2855}. In the following section, we explore alternative strategies to reduce both the computational cost and the hardware connectivity demands of the computation of $N(q)$ and $\epsilon(q)$, aiming to enhance its near-term feasibility.

\section{Measuring $N(q)$  and $\varepsilon(q)$  without preparing the single-spinon state\label{sec:norm-disp}}

The direct implementation of the  $\ket{\Psi(q)}$ on quantum computers necessitates ancillary qubits, leading to a non-deterministic algorithm \cite{nielsen2010quantum}. In this section, we explore alternative approaches to reduce the hardware requirements and compute both the norm $N(q)$ and the energy $\epsilon(q)$ by parts. We consider two different methodologies: a direct computation through expectation values of certain operators in the ground state wave function and their estimation using the Hadamard test.

\
\subsection{Expectation value of strings of SWAP Operators}
Both the norm and energy can be evaluated through expectation values. We can write the norm as
\begin{equation}
    N(q)=\frac{1}{L+1} \sum_{m=0}^L \sum_{n=0}^L e^{iq(m-n)} \bra{\Psi(n)}\ket{\Psi(m)},
\label{parts}
\end{equation}
being $\ket{\Psi(n)}=SWAP^{n}_{0}\ket{\Psi(0)}$. Therefore, we can obtain the norm by computing the overlaps with a quantum computer as expectation values 
\begin{equation}    \bra{\Psi(n)}\ket{\Psi(m)}=\bra{\Psi(0)}({\rm SWAP}^n_{0})^{\dagger}{\rm SWAP}^m_{0}\ket{\Psi(0)}
\end{equation}
where $\ket{\Psi(0)}$ is obtained trivially once we have the ground state,
and then, reconstructing classically the sum in \eqref{parts}. However, while the expectation value of a single permutation $\bra{\Psi(m+1)}\ket{\Psi(m)}$ involves only four Pauli terms, the number of Pauli observables grows as $4^p$, being $p=|n-m|$. For instance, for $p=2$ ($n=2,m=0$) we have 16 Pauli strings:
\begin{equation}
\begin{split}
{\rm SWAP}_0^2&= {\rm SWAP}^2_1 {\rm SWAP}^1_0\\
&=\frac{1}{4}(III+XXI+YYI+ZZI \\
&+IXX+XIX+iYZX-iZYX\\
&+IYY-iXZY+YIY+iZXY\\
&+IZZ+iXYZ-iYXZ+ZIZ).\\
\end{split}
\end{equation}
Given that the maximum value of $p$ is $L$, and that the observables in ${\rm SWAP}_0^n$ are contained within ${\rm SWAP}_0^{n+1}$, the total number of observables is $\mathcal{O}(4^L)$, which scales exponentially with the number of qubits. 

Analogously, a similar argument applies to the energy, 
\begin{eqnarray} 
    \bra{\Psi(q)}H_{L+1} \ket{\Psi(q)} =\nonumber\\
    \frac{1}{L+1} \sum_{m=0}^L \sum_{n=0}^L e^{iq(m-n)} \bra{\Psi(n)}H_{L+1}\ket{\Psi(m)},
\label{eq:parts2}
\end{eqnarray}
We have not attempted a specific counting of the number of Pauli terms needed to compute \eqref{eq:parts2} that will depend, of course, on the Hamiltonian. In principle, this can be performed via symbolic computation. However,  avoiding an exponential cost appears difficult and may be restricted to particular cases. Such limitations are common across other existing approaches to simulate quasiparticles, as exemplified in Ref.~\cite{Balents}. In summary, here the quantum circuit only needs $L+1$ qubits without ancillary, but incurs an exponential number of terms to compute. This is to be compared with the approach in the previous section, which required twice as many qubits ($2L+1$  vs $L+1$) qubits,   and a polynomial number of Fredkin gates, along with the additional SWAP operations conditioned by the hardware connectivity.

\subsection{Hadamard test}
\begin{figure}[t]
\centering
    \includegraphics[width=0.9\columnwidth]{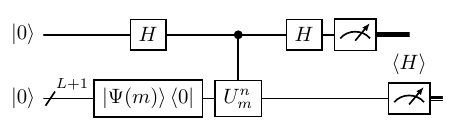}
    \caption{Hadamard test to estimate the real part of the overlaps $\langle{\Psi(m)}|\Psi(n)\rangle$ and transition amplitudes $\langle{\Psi(m)}|H|\Psi(n)\rangle$.}
    
\label{fig:hadamard}
\end{figure}

We now introduce a third approach that makes use of 
the Hadamard test \cite{Hadamard}. The scheme of the quantum circuit, shown in Fig.\,\ref{fig:hadamard}, uses a single ancillary qubit. The Hadamard test works as follows.  First, we initialize in an equal superposition with a Hadamard gate:
\begin{equation}
    \ket{\Phi}=\frac{1}{\sqrt{2}}[\ket
{0}\otimes\ket{\Psi(m)}+\ket{1}\otimes \ket{\Psi(m)}]
\end{equation}
We then apply a controlled operation $U_m^n$
\begin{equation}
\begin{split}
    \ket{\Phi}&=\frac{1}{\sqrt{2}}[\ket{0}\otimes\ket{\Psi(m)}+\ket{1}\otimes U_m^n\ket{\Psi(m)}]\\    
    &=\frac{1}{\sqrt{2}}[\ket{0}\otimes\ket{\Psi(m)}+\ket{1}\otimes \ket{\Psi(n)}],
\end{split}
\end{equation}
which implements the unitary operation ${\rm SWAP}_m^n$, a product of ${\rm SWAP}$ gates which move the spin from $m$ to $n$. That is,
\begin{equation}
    U_m^n=\mathrm{SWAP}_{n-1}^n \cdots \mathrm{SWAP}_{m}^{m+1} \quad (n > m)
\end{equation}

To complete the protocol, we apply again a Hadamard gate in the ancillary qubit

\begin{equation}
    \ket{\Phi}=\frac{1}{2}[\ket
{0}\otimes(\ket{\Psi(m)}+\ket{\Psi(n)}+\ket{1}\otimes (\ket{\Psi(m)}-\ket{\Psi(n)}],
\end{equation}
and after measuring we find that the difference probability $p_0-p_1$ between the two possible outputs is $Re \bra{\Psi(m)}\ket{\Psi(n)}$, or $Im \bra{\Psi(m)}\ket{\Psi(n)}$ if we introduce an $S^\dagger$ before the last Hadamard gate \cite{PhysRevResearch.5.043087}. As $N(q)$ is real, we can estimate the norm as
\begin{equation}
    N(q)=\frac{2}{L+1} \sum_{n\geq m}^{L} \cos(q(m-n)) \bra{\Psi(n)}\ket{\Psi(m)}.
\end{equation}
The cost of estimating $N(q)$ comes from the number of quantum circuits to execute,

\begin{equation}
   \frac{(L)(L+1)}{2},
\end{equation}
where $L+1$ circuits in the sum do not need to be evaluated as $\bra{\Psi(m)}\ket{\Psi(m)}=1$. 

To compute the $\bra{\Psi(q)}H_{L+1}\ket{\Psi(q)}$, one just needs to measure in the corresponding basis the remaining qubits. Measuring the ancillary qubit collapses the system into one of the following superposition states:  
\begin{equation*}
\begin{split}
\ket{\Psi_+^{mn}} &= \frac{1}{\sqrt{2}} \left( \ket{\Psi(m)} + \ket{\Psi(n)} \right) \quad \text{if the outcome is } 0, \\
\ket{\Psi_-^{mn}} &= \frac{1}{\sqrt{2}} \left( \ket{\Psi(m)} - \ket{\Psi(n)} \right) \quad \text{if the outcome is } 1.
\end{split}
\end{equation*}

Then, by measuring the computational qubits we can obtain $\bra{\Psi_+^{mn}}H_{L+1}\ket{\Psi_+^{mn}}$ and $\bra{\Psi_-^{mn}}H_{L+1}\ket{\Psi_-^{mn}}$. The difference between these expectation values allows for the estimation of the real and imaginary components of the transition amplitudes 

\begin{equation}
  t_{mn}=\bra{\Psi(m)}H_{L+1}\ket{\Psi(n)}
\end{equation}
simultaneously with the overlaps $\bra{\Psi(m)}\ket{\Psi(n)})$. Assuming that the Hamiltonian can be decomposed into \( N_g \) groups of mutually commuting terms, enabling joint measurement, this results in a total of
\begin{equation}
    N_g(L+1)(L+2)/2
\end{equation}
circuits, where $L+1$ circuits correspond to the case $n=m$ that can be computed without the ancillary qubit. This procedure imposes fewer hardware requirements than the LCU approach, requiring a single ancilla qubit and $\mathcal{O}(L)$ controlled operations, making it particularly well-suited for near-term quantum devices with limited qubit connectivity and coherence.

The Hadamard test approach
can be regarded as the minimal and most natural realization that generalizes to the LCU structure depicted in Fig.\,\ref{fig:LCU}. The LCU approach enables the simultaneous probabilistic evaluation of all components by directly preparing the state $\ket{\Psi(q)}$. In contrast, the present procedure provides a deterministic, term-by-term estimation of the norm and energy contributions, at the cost of executing more circuits. The resource costs of all methods are summarized in Table~\ref{tab:overlap_methods}. 

\begin{table}[h]
\centering
\begin{tabular}{l c c c}
\hline\hline
Method & Ancilla qubits & Fredkin gates & Circuits \\
\hline
LCU             & $L$ & $\mathcal{O}(L^2)$ & $\mathcal{O}(N_g)$\textsuperscript{\textdagger} \\
SWAP operator   & $0$ & $0$ & $\mathcal{O}(4^L)$\textsuperscript{\textdaggerdbl} \\
Hadamard test   & $1$ & $\mathcal{O}(L)$ & $\mathcal{O}(N_g L^2)$ \\
\hline\hline
\multicolumn{4}{l}{\textsuperscript{\textdagger}\footnotesize{Post-selection on $\ket{0}^{\otimes L}$ succeeds with probability $ N(q)/(L+1)$.}}\\
\multicolumn{4}{l}{\textsuperscript{\textdaggerdbl}\footnotesize\parbox[t]{0.95\linewidth}{%
\raggedright This scaling can be improved using commuting-grouping strategies \cite{TILLY20221,Huang2020}. In the case of the energy, it will depend on the specific form of the Hamiltonian.}}\\
\end{tabular}
\caption{Resource scaling for selected quantum norm and energy estimation methods.}
\label{tab:overlap_methods}
\end{table}

\begin{figure*}[t]

\includegraphics[scale=0.95]{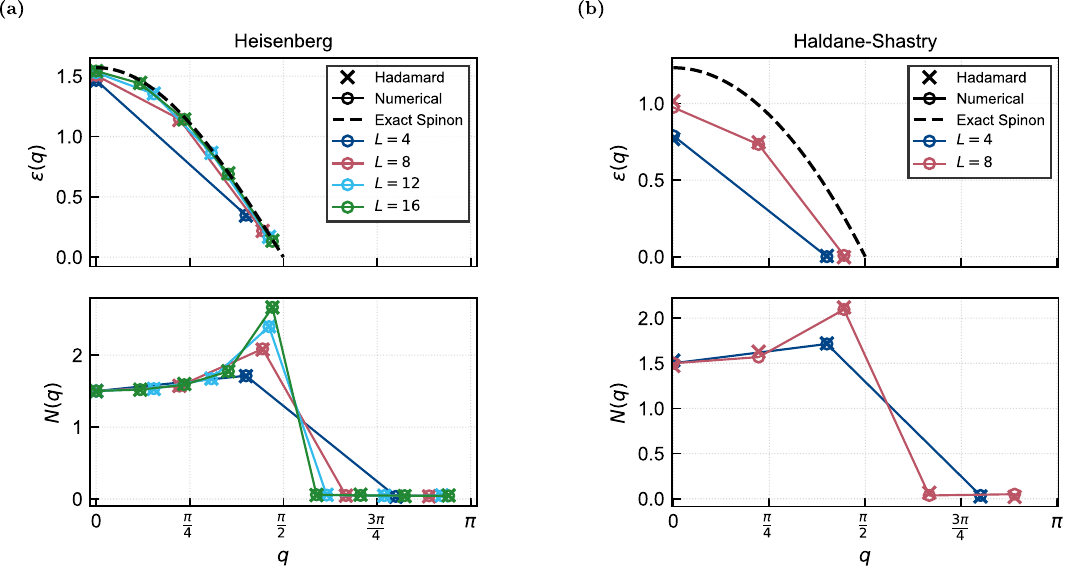}~~~~
\caption{Comparison between the energy dispersion $\epsilon(q)$ and norm $N(q)$ estimated via the Hadamard test (with $N_s = 10^4$ shots) and exact numerical simulations for the one-dimensional Heisenberg antiferromagnet (a) and Haldane-Shastry (b) chains. The ground states of both models were prepared using VQE (Heisenberg) and the Gutzwiller projection (Haldane-Shastry), as outlined in Sec.~\ref{sec:GSPREP}. Cross markers denote results from the Hadamard test, while open circles connected by solid lines   represent numerical data obtained via exact diagonalization and following Ref.\,\cite{kulk}. Excellent agreement is observed between the quantum and numerical approaches, with closer correspondence for the Heisenberg chain as $L$ increases. This discrepancy arises from the non-deterministic state preparation in the Haldane-Shastry model, which exponentially decreases the number of successful protocol executions for estimating $\epsilon(q)$ and $N(q)$. Consequently, achieving comparable precision in the Haldane-Shastry chain requires an exponential increase in the number of shots with system size, as the success probability of ground-state preparation decreases exponentially.}
\label{Hadamard}
\end{figure*}

\subsection{Estimation of the single-spinon energy}
In Fig.\,\ref{Hadamard}, we show the results of the estimation of the single-spinon energy (Eq.\,\eqref{eq:spinon-disp}) and the norm $\bra{\Psi(q)}\ket{\Psi(q)}$, for the Heisenberg (Haldane-Shastry) model, for values of $L=16$ ($L=8$), using in-silico simulations of the Hadamard test approach outlined in the previous subsection. In the case of the Heisenberg model, the ground state is prepared using VQE, whereas in the Haldane-Shastry model, we are using the Gutzwiller approach.  In both cases, we take $L$  even, so that the ground state of the parent chain has $S=0$, and we also choose cases where $L/2$ is an even integer, following Kulka \cite{kulk}. See Appendix A for the discussion of the results for odd $L/2$. 

For both models, we compare the results of the simulated quantum computing approach with those of in-silico calculations.  We find an excellent agreement. We also include a dashed black line, the analytical single-spinon dispersion. The role of finite-size effects is apparent. As we increase $L$, our results get closer to the analytical results (see also Fig. \ref{fig:HS}). Nevertheless, by choosing the non-deterministic preparation of the Haldane-Shastry ground state, all of the previously discussed methods will require an increase in the number of shots to offset the preparation failure rate and maintain constant standard deviations. This can be seen in Fig.\,\ref{Hadamard}, where the results for the Haldane-Shastry model exhibit increasing deviations with  $L$, in contrast to those of the Heisenberg chain.

\section{Summary and Conclusions}
Spinons were proposed \cite{FADDEEV1981375} as the hidden  $S=1/2$ elementary particles, with a well-defined energy-momentum relation $E(q)$ that, when glued together,   make the physical excitations of certain spin models, such as the one-dimensional $S=1/2$ antiferromagnetic Heisenberg spin chain with first neighbour coupling or the Haldane-Shastry model. Defining a single-spinon wave function has been particularly challenging \cite{talstra97}. In a recent work, Kulka and coworkers \cite{kulk}  proposed a heuristic wave function for single spinons associated with spin chains with $L$ sites, as an object living in chains with $L+1$ sites, with a $S=1/2$ wave function determined by the ground state of the spin chain with $L$ sites. A useful property of their ansatz is that it permits one to retrieve both the single-spinon dispersion energy, as well as the domain of existence in the reciprocal space.

The main merit of the single-spinon ansatz is to provide a 
phenomenological route to describe single spinons, and even to connect them to experiments \cite{zhao25}. This paper is devoted to pave the way towards the future 
implementation of the recently proposed \cite{kulk} single-spinon ansatz in a quantum computer. Given that the single-spinon wave function is not an eigenstate of a Hamiltonian, quantum computers may be the best way to flesh out single-spinons.  In addition, our work serves to assess whether quantum computers could be used to evaluate the single-spinon dispersion energy and domain of existence in reciprocal space.

The main results of this work are the following:
\begin{itemize}[leftmargin=*]
\item We answered in the positive the question of whether the single-spinon ansatz \cite{kulk} can be used to compute the single-spinon dispersion of the Haldane-Shastry Hamiltonian. Hence, this extends to the case of models with long-range exchange the range of validity of the original paper of Kulka {\em et al.}~\cite{kulk}.
\item We have proposed a non-deterministic quantum algorithm to prepare the single spinon ansatz in the case of two  $S=1/2$ one-dimensional models with periodic boundary conditions, namely, the antiferromagnetic Heisenberg chain and the Haldane-Shastry Hamiltonian. The algorithm has two main steps: the preparation of the ground state of the model and then the application of a linear combination of unitaries, which is non-deterministic.
In principle, this preparation could be exploited to obtain the single spinon energy dispersion. 
\item We have proposed an approach to prepare the ground state of the Haldane-Shastry model in a quantum computer, taking advantage of two known results: first, that this state can be obtained from the Gutzwiller state for the first-neighbour Hubbard model; second, the existence of an algorithm to prepare the Gutzwiller state in quantum computers \cite{Murta21}.

\item We have proposed two complementary methods to determine the single-spinon energy dispersion, exploiting the single-spinon ansatz, but  without the need to prepare the single-spinon state in a quantum computer

\end{itemize}

In summary,  we have paved the way towards 
the implementation of the recently proposed \cite{kulk} single-spinon ansatz in a quantum computer. On the theory side,  future work could address the effect of boundary conditions, studying the properties of extensions of the single-spinon ansatz to open-end chains, both in silicon and in quantum processors.  Other properties of the single-spinon ansatz, such as spin correlators and entanglement, could be explored. The extension of the single-spinon concept to more complicated one-dimensional spin Hamiltonians \cite{shastry81} or even  
to two-dimensional models \cite{tang13} seems also a very interesting venue.

\begin{acknowledgments}
D. Faílde and A. Gómez were supported by MICINN through the European Union NextGenerationEU recovery plan (PRTR-C17.I1),
and by the Galician Regional Government through the “Planes
Complementarios de I+D+I con las Comunidades Autonomas”
in Quantum Communication.  Simulations on this work were performed using the Galicia Supercomputing Center (CESGA) FinisTerrae III supercomputer with
financing from the Programa Operativo Plurirregional
de España 2014-2020 of ERDF, ICTS-2019-02-CESGA3, and the Qmio quantum infrastructure, with financing
from the European Union, through the Programa Operativo Galicia 2014-2020 of ERDF REACT EU, as part
of the European Union’s response to the COVID-19 pandemic. J.F.-R. acknowledges financial support from  SNF Sinergia (Grant Pimag), FCT (Grant No. PTDC/FIS-MAC/2045/2021),  from Generalitat Valenciana (Prometeo2021/017 and MFA/2022/045), and MICIN-Spain (Grants No.PID2022-141712NB-C22 and PRTR-C17.I1).

\end{acknowledgments}

\appendix
\renewcommand{\thefigure}{\thesection.\arabic{figure}}
\setcounter{figure}{0}

\section{Numerical Details and Parity Effects in the Single-Spinon Ansatz}

\begin{figure}[h]
\centering
\includegraphics[scale=1.2]{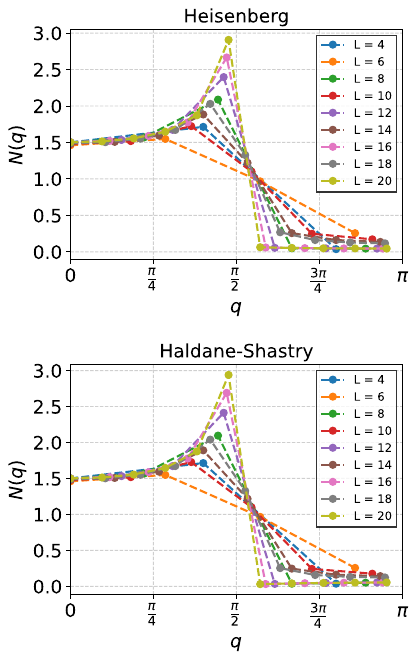}~~~~
\caption{Norm $N(q)$ for the Heisenberg and Haldane-Shastry models, $L = 4$ to $20$.}
\label{L_odd}
\end{figure}

Throughout this work, we have presented results for the single-spinon ansatz on chains with $L/2$ even. However, the norm $N(q)$ of the state $\ket{\Psi(q)}$ exhibits a slightly distinct behaviour depending on the parity of $L/2$. Following the procedure outlined in Ref.\,\cite{kulk}, chains with $L/2$ odd do not display the same strict vanishing behaviour in $N(q)$ as those with $L/2$ even. As illustrated in Fig.~\ref{L_odd}, the norm for $L/2$ odd chains is larger than for $L/2$ even chains. Nevertheless, numerical results indicate that the norm for $L/2$ odd chains decreases with $L$ in the region $q>\pi/2$, consistent with a vanishing norm in the thermodynamic limit.  Although this behaviour does not have an apparent explanation, it should be considered carefully, as the vanishing of the norm $N(q)$, which signals an unphysical state, is precisely what allows discarding the energy spectrum for $q > \pi/2$.

Furthermore, as shown in Fig.~\ref{Hadamard}, our results for $\epsilon(q)$ in the Heisenberg model with $L=16$ differ from those reported in Ref.~\cite{kulk}. As we obtain the same norm, this discrepancy needs to arise from the value of the ground state energy $E_0^{L+1}$ used in \eqref{eq:spinon-disp}. 
In our calculations, we obtained $E_0^{L+1}$ from exact diagonalization of the Hamiltonian on a chain of $L+1$ sites. 
No specific details regarding the computation or choice of this energy are provided in Ref.~\cite{kulk}, which makes a direct comparison difficult. 

\begin{figure}

\includegraphics[width=0.45\textwidth]{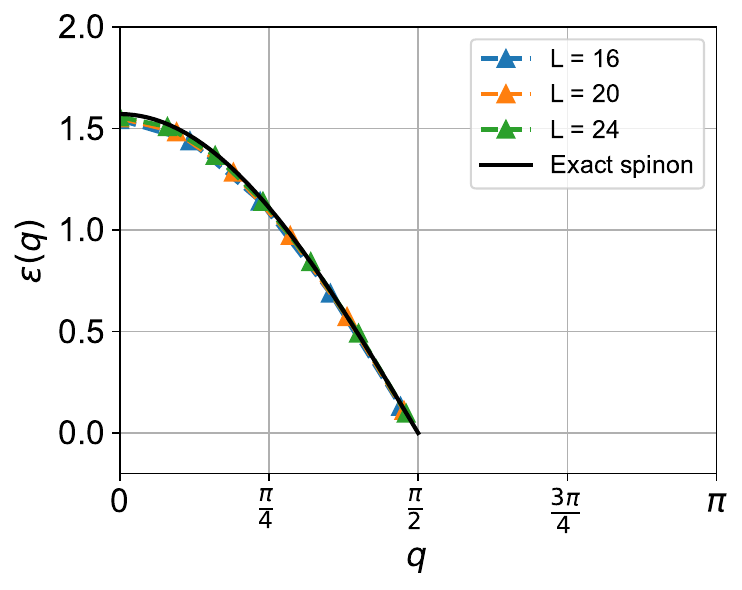}~~~~
\caption{Energy dispersion $\epsilon(q)$ for the Heisenberg model, estimated using $E_0^{L+1} = J(L+1) \left( \frac{1}{4} - \ln 2 \right)$ for system sizes $L =\lbrace{ 16,20,24\rbrace}$.}
\label{ku}
\end{figure}

Using the value of $E_0^{L+1}$ obtained by diagonalizing the Hamiltonian numerically is only possible for moderate system sizes, but not for large $L$. For the Heisenberg model, which is exactly solvable via the Bethe ansatz, an alternative for large chains is to use the following expression that arises in the thermodynamic limit \cite{Bethe}:
\begin{equation}
    E_0^{L+1} = J(L+1) \left( \frac{1}{4} - \ln 2 \right).
\end{equation}
This results in a dispersion relation $\epsilon(q)$ that is in closer agreement with the one reported in Ref.~\cite{kulk} (Fig.\,\ref{ku}).

However, for exploring spinons in higher-dimensional systems or non-analytically solvable models, this is not a crucial element of the procedure as it corresponds to a global energy shift that does not affect the shape of the dispersion relation. The unique motivation here to take it into account is to properly compare our results with the theoretical spinon dispersion relation for the Heisenberg and Haldane-Shastry models.

\section{Free fermions as initial state of the 1D Heisenberg and Haldane-Shastry models}\label{B}

The VQE approach offers an approximate but deterministic route to ground-state preparation in the Heisenberg Haldane-Shastry model. This contrasts with the exact but probabilistic methods based on the Bethe ansatz and Gutzwiller projection applied to the free-fermion Hamiltonian with nearest-neighbor interactions. Thus, a trade-off arises between accuracy and computational cost, which must be carefully weighed. 

Choosing the VQE approach entails the well-known challenges of variational optimization, such as barren plateaus and convergence to local minima. For that matter, the initialization of the quantum circuit in an easy-to-prepare state close to the ground state (warm start) is a crucial step to facilitate convergence and avoid traps during the optimization. For the Heisenberg and Haldane-Shastry models, the $H_{XY}$ ground state, $\ket{\Psi_{XY}}$, provides a natural starting point, capturing much of the dominant contributions in each Hamiltonian.

\begin{figure}[t] 
    \centering 
    \hspace{-0.8cm}\includegraphics[width=0.45\textwidth]{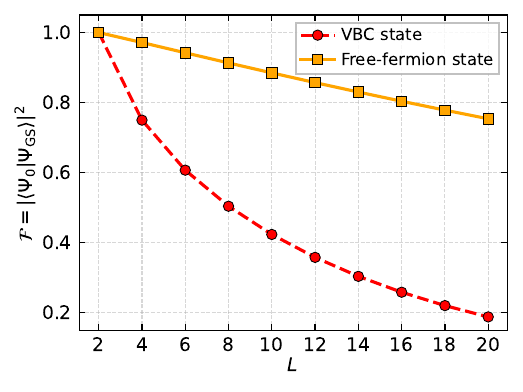} 
    \caption{Fidelity $\mathcal{F}=|\bra{\Psi_0}\ket{\Psi_{GS}}|^2$ between the initial state $\ket{\Psi_0}$ and the ground state of the 1D antiferromagnetic Heisenberg model $\ket{\Psi_{GS}}$ as a function of the number of spins $L$.} 
    \label{fig:fidelidad} 
\end{figure}

For a Heisenberg chain of $L$ sites, initializing in $\ket{\Psi_{XY}}$ with $O(L)$ Givens rotations significantly enhances the overlap with the true ground state compared to a constant depth VBC state (Fig.\,\ref{fig:fidelidad}). This allows the low infidelities reported to be maintained for medium-length chains with a modest number of parameters and layers.

\begin{figure}[h] 
    \centering 

    \hspace{-0.8cm}\includegraphics[width=0.45\textwidth]{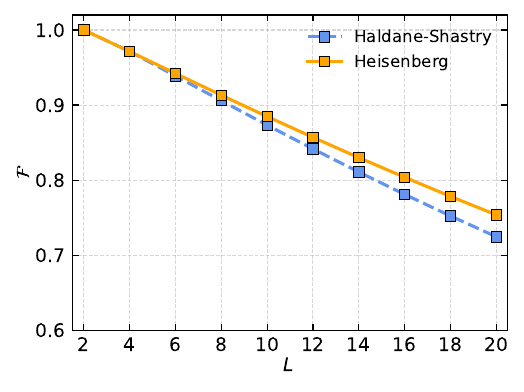} 
    \caption{ Fidelity  between the exact ground state (obtained numerically) of the  Heisenberg and Haldane-Shastry models and the ground state of the XY model, as a function of chain length $L$.} 
    \label{fig:FF} 
\end{figure}

The Haldane-Shastry Hamiltonian introduces further challenges: its all-to-all interactions require circuits with high connectivity and longer gate sequences.
Nevertheless, as in the Heisenberg chain, the ground state of $H_{XY}$, $\ket{\Psi_{XY}}$, remains a suitable initial state for the variational ansatz, as illustrated in Figure~\ref{fig:FF}. This makes it not only valuable as an initial state for tailored variational approaches but also for other algorithms such as Quantum Phase Estimation \cite{Hadamard} applied to medium-length chains (Fig.\,\ref{fig:FF}).

\subsection{Variational preparation of the Haldane-Shastry ground state via VQE}

We conclude by presenting numerical results for the variational preparation of the Haldane–Shastry ground state. For the quantum circuit, we again employ a HVA-inspired ansatz, where each block is constructed from the Pauli operators corresponding to interactions of range $\kappa$. That is, we prepare the following variational wave function

\begin{equation}
    \ket{\Psi(\boldsymbol{\theta)}}= \prod_j^l \prod_{\kappa_{max}}^1 U_{\kappa,j} (\vec{\theta}_{\kappa,j})\ket{\Psi_{XY}},
\end{equation}
where $\kappa$ denotes the interaction range, goes from $\kappa_{\max}$ to $1$ (nearest neighbors) considering periodic boundary conditions, $j$ indexes the layer, and  

\begin{equation}
    U_{\kappa,j}(\boldsymbol{\theta}_{\kappa,j}) = \prod_{s} e^{-i \theta_{s}^{\kappa,j} H_{s,\kappa}}
\end{equation}
is a product of unitaries built from mutually commuting terms of $H_\kappa$. The initial state $\ket{\Psi_{XY}}$ is the free-fermion ground state (highlighted in yellow in Fig.~\ref{fig:infidelity_error}a), while the final block in each layer corresponds to the nearest-neighbor parameterized ansatz (shown in green in Fig.~\ref{fig:infidelity_error}a). The number of layers was set to $L/2$ in accordance with the simulations performed in Section \ref{subsec:1DHEIS}. The remaining optimization details are the same as in the VQE implementation of the Heisenberg model.

\begin{figure}[h] 
    \centering 

    \hspace{-0.8cm}\includegraphics[width=0.475\textwidth]{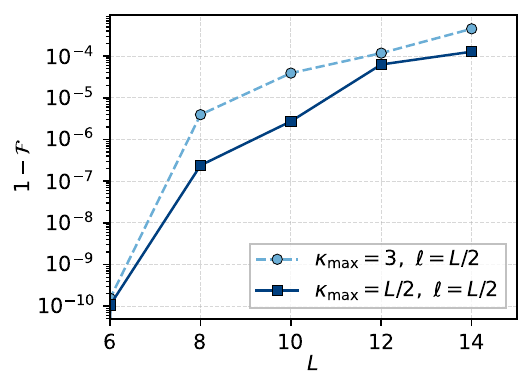} 
    \caption{
    Average fidelity over ten independent experiments between the exact ground state (obtained numerically) of the Haldane--Shastry model and the variational state optimized via energy minimization.  The solid blue curve shows results for an unrestricted variational ansatz with all-to-all interactions, while the dashed curve contains the results after restricting the quantum circuit to a maximum range of three nearest neighbors (\emph{i.e.}, $\kappa \le 3$).
    }
    \label{fig:VQE_Haldane} 
\end{figure}

In Figure~\ref{fig:VQE_Haldane}, we plot the average infidelities obtained after ten VQE executions with different parameter initializations, for spin chain lengths $L \in \{6, 8, 10, 12, 14\}$. We consider two variational ans\"atze: (i) one with all-to-all connectivity ($\kappa_{\max} = L/2$), and (ii) one restricting interactions to at most three nearest neighbors ($\kappa_{max} = 3$). This restriction is motivated by the quadratic decay of interaction strength with distance in the Haldane-Shastry model. Within the range of system sizes studied, the ansatz exploiting long-range interactions yields slightly lower infidelities than the short-range-restricted variant. However, this improvement comes at the cost of a larger number of variational parameters and increased circuit depth, rendering the fully connected ansatz more susceptible to barren plateaus and noise on near-term quantum hardware. On this matter, a more thorough analysis, including the impact of shot-noise effects and experimental validation on real hardware is required.

\end{document}